# Tablet-based Information System for Commercial Aircraft: Onboard Context-Sensitive Information System (OCSIS)

Wei Tan[1] and Guy A. Boy[2][1]

[1] School of Human-Centered Design, Innovation and Art, Florida Institute of Technology, FL 32901, USA
[2] ESTIA / Air and Space Academy, F 64210 Bidart, France
[1] weitan2011@outlook.com
[2] guy.andre.boy@gmail.com

**Abstract.** Pilots currently use paper-based documentation and electronic systems to help them perform procedures to ensure safety, efficiency and comfort on commercial aircrafts. Management of interconnections among paper-based operational documents can be a challenge for pilots, especially when time pressure is high in normal, abnormal, and emergency situations. This dissertation is a contribution to the design of an Onboard Context-Sensitive Information System (OCSIS), which was developed on a tablet. The claim is that the use of contextual information facilitates access to appropriate operational content at the right time either automatically or on demand. OCSIS was tested using human-in-the-loop simulations that involved professional pilots in the Airbus 320 cockpit simulator. First results are encouraging that show OCSIS can be usable and useful for operational information access. More specifically, context-sensitivity contributes to simplify this access (i.e., appropriate operational information is provided at the right time in the right format. In addition, OCSIS provides other features that paper-based documents do not have, such as procedure execution status after an interruption. Also, the fact that several calculations are automatically done by OCSIS tends to decrease the pilot's task demand.

**Keywords:** Commercial Aircraft, Onboard Information System, Human-Centered Design, Tangible Interactive System, Avionics, Context.

## 1 Introduction

An airplane consists of a number of mechanical and computerized systems. An airplane cannot stop or brake in the air, and fuel is consumed during the entire flight. Consequently, flight time is limited. Flight crewmembers have all the capacities and limitations of any human being; they can be qualified as human operators. They typically collaborate, communicate, and cooperate with each other to execute flight tasks. Actions performed by flight crewmembers in the cockpit must adhere to procedures in context. Onboard paper-based operational documents barely provide context-sensitive information. Therefore, context has to be handled by pilots.

---

[1] This work was done while Dr. Boy was Dean and University Professor at Florida Institute of Technology.

Procedures execution is a major safety factor. Until now, specific pilot roles have been supported by paper-based documentation in both operations and training. It is a pilot's job to make decisions, act, communicate, cooperate, and coordinate operationally, with procedures established in operational documentation developed through airline policy and governmental regulation. All the procedures and information can be found from documents. The onboard documents can be categorized into four kinds of documents: flying documents, which are related to all flight operations; systems documents, which include systems' theory, principles, and controls; navigation documents, which are the charts that pilots use on the flight deck; and performance documents, which provide operational data for all flight phases such as takeoff, landing, and go-around [1].

However, onboard paper-based documents are not the only resources that pilots have in the cockpit. Several other onboard systems can enhance the pilot's awareness of aircraft status (i.e., aircraft states). They can provide very comprehensive information on the state of the aircraft in an integrated way. Taking Airbus Electronic Centralized Aircraft Monitor (ECAM) as an example, it provides actions together with corresponding flight parameters to pilots who have to deal with related malfunctions. It provides steps to handle failures for a large number of situations [2].

## 2 State of art

### 2.1 Tablet-based systems onboard

Pilots are familiar with paper-based manuals, which are easy to use, tag, mark, and retain, even though they are heavy and difficult to carry. Nobody can permanently remember all procedures and technical knowledge, particularly, under time pressure. Now many applications on iPad that contain paper charts information are available. Moreover, Boeing introduced a tablet-based version of the paper Quick Reference Handbook (QRH) used by flight crews in 2013 [3]. The Interactive QRH offers advanced navigation and search capabilities to enable the pilot to easily find the proper checklist (see Fig. 1). The Interactive QRH also simplifies non-normal checklist use, especially for those checklists in which the correct condition must be selected from two or more choices. In line with technological innovation, Airbus has developed iPad EFB solutions to provide airlines with an alternative to PC operating system EFB devices (see Fig. 1). With "FlySmart with Airbus" applications on iPad, pilots will be able to compute performance calculations and also consult Airbus Flight Operations Manuals from a light hand-held device [4, 5].

Even though Electronic Flight Bag (EFB) and Onboard Information Systems (OIS) are advanced tools to assist pilots' work, operational documents are still in their original format and arrangement. Not all abnormal procedures are available on the ECAM, nor do all types of aircrafts have ECAM or some other electronic systems to process and display procedures. Unlike paper, computer support enables easy contextualiza-tion of provided information. The Onboard Context-Sensitive Information System (OCSIS) is introduced into commercial aircraft cockpit to make the flight safe, efficient, and comfortable by providing assistance in normal and abnormal situations and enhanced capabilities of interaction with other onboard systems.

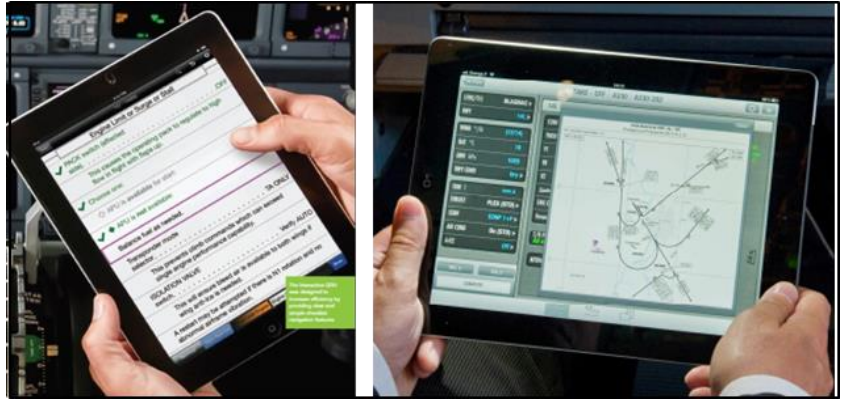

**Fig. 1.** Interactive QRH in B737NG [3] on the left and Airbus FlySmart [5] on the right

### 2.2 Human-centered design approach

"A human-in-the-loop (HITL) simulation is a modeling framework that requires human interaction. This approach is typically called participatory design. The emergence of HITLS technologies, therefore, enables researchers and practitioners to investigate the complexity of human-involved interactions from a holistic, systems perspective [6]".

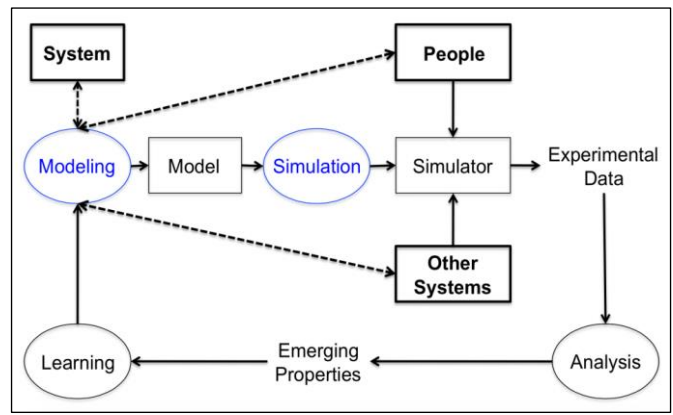

**Fig. 2.** Human-Centered Design approach [7]

As shown in Fig.2, the model typically represents reality in a simplified way. It proposes important elements and their relevant interconnections in an appropriate, orchestrated manner. The simulation represents the interaction that brings the model to life, which can be used to improve understanding of interactions among different elements that the model implements. It is also used to improve the model itself and eventually modify it [7]. HITLS is used early on during the design phase. Consequently, Human-Centered Design (HCD) has been used to incrementally improve OCSIS toward an acceptable mature version (i.e., incremental prototype development, test, and modification) [8]. Modeling OCSIS requires pilots' involvement, and interaction with other onboard systems. The process can be run on a flight simulator, which in turn produces experimental data that could be used to improve OCSIS.

### 2.3 Context-sensitive procedures

"Context is any information that can be used to characterize the situation of an entity. An entity is a person, place, or object that is considered relevant to the interaction between a user and an application, including the user and applications themselves [9]". Pilots must accomplish flight tasks (e.g., cockpit preparation, takeoff, approach, and landing procedures) within the appropriate context depend on flight phases. We use Interaction Blocks to represent, implement, and handle context-sensitive procedures (see Fig. 3). Fig. 4 provides an example of iBlock for flaps setting.

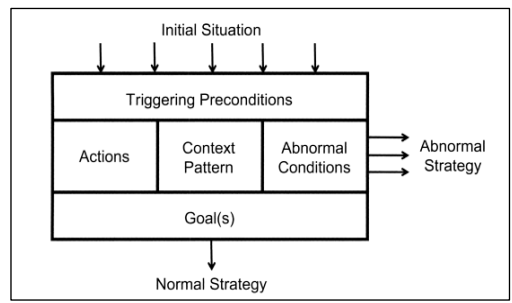

**Fig. 3.** Interaction block representation [10]

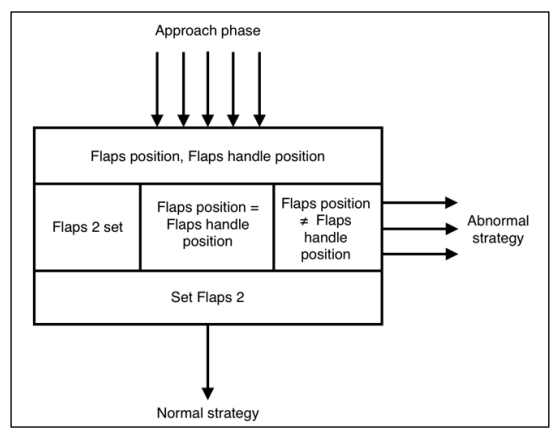

**Fig. 4.** The iBlock for "Flaps Locked"

An interaction block is defined by: a set of actions; and a situation pattern that includes triggering preconditions and a context pattern; and post-conditions that include a goal and abnormal conditions [10]. Let's take the example of the "Flaps Set" procedure in the approach scenario (see Fig. 4). Its triggering preconditions consist of "Flaps position" (i.e., visible on the E/WD screen) and "Flaps handle position" (i.e., visible on the Flaps lever). When the "Flaps position" on E/WD equals to the value of the "Flaps handle position," then the goal is reached and the pilot can continue to next procedure; otherwise, a pop-up window is displayed on OCSIS to inform about an abnormal situation, and the "Flaps Locked" procedure needs to be executed. This procedural knowledge representation was developed during the 1990s to represent operational procedures in aircraft cockpits and led to deeper investigations on context representation also. It is therefore very appropriate for context-sensitive procedural information representation.

## 3 Design of OCSIS prototype

OCSIS is currently programed using Objective-C, an object-oriented language available on Apple's hardware, on Xcode, the Integrated Development Environment for Objective-C. Human-Centered Design institute (HCDi) is equipped with two commercial aircraft simulators, A320 and B737, that were developed using Prepare3d, a Lockheed Martin software, which allows vehicle control (e.g., marine, terrestrial, air, or spatial) in a totally virtual world that is simulated to be close to reality.

The first prototype of OCSIS is applying A320 procedures and references. Once the application starts on the iPad, a Welcome page is displayed. The default/initial page displays procedures and actions that crews need to perform or have performed. A menu is provided to select other tabs at the top of this page (see Fig. 5).

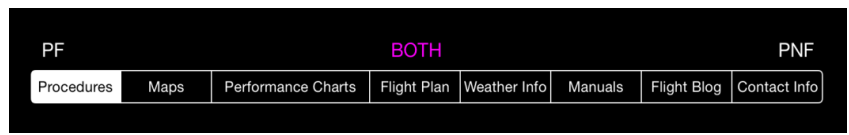

**Fig. 5.** OCSIS's Menu

At the end of the 1990s, EURISCO and Airbus carried out a study with 60 commercial pilots on how electronic documentation could be structured [11]. Results showed that it could be best structured into three information levels. We adopted this re-organized structure for OCSIS (see Fig. 6):

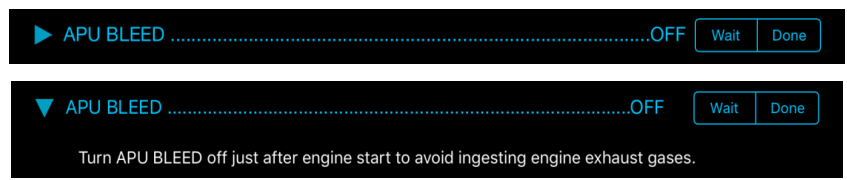

**Fig. 6.** Level 1 and Level 2 information of APU Bleed set

Level 1: Need to know or safety-critical information that the pilot needs to have immediately.

Level 2: Nice to know or short explanations of Level 1.

Level 3: Understand technical knowledge on systems' principles and trouble-shooting.

In order to keep consistency of Airbus's philosophy and make all the actions are easy to be understand, Dynamic Color System (DCS) is designed to enhance the pilot's perception and comprehension of the current situation., as shown in Table 1. Different colors stand for different meanings that provide the pilot with a direct and swift status of procedures. The items marked with "*" are updated after a series of formative evaluations.

**Table 1.** Color codes

| Color | Representation |
|---|---|
| Cyan | Actions to be performed |
| Green | Actions performed |
|  | Marked as performed |
| Amber | Postponed actions or checks |
|  | *The title of abnormal procedures |
|  | Cautions |
| Red | *The title of emergency procedures |
|  | Warnings |
| White | Notes |
|  | More information for actions |
|  | *The title of flight phase |
|  | *Normal Checklists |
| Magenta | Restrictions or constraints |
| Grey | *Not applicable for current context |

"Ready to do" actions are in cyan. Once the action is completed and OCSIS can access the related parameters' status, it automatically becomes "green." In the current version of OCSIS, this kind of automation is done for only the a few parameters, which can be detected (i.e., colors change automatically). For all other parameters, the pilot marks them "DONE" manually (i.e., the action line becomes green). The pilot can also postpone an action by selecting the "WAIT" button, and then the action line becomes amber (see Fig. 7).

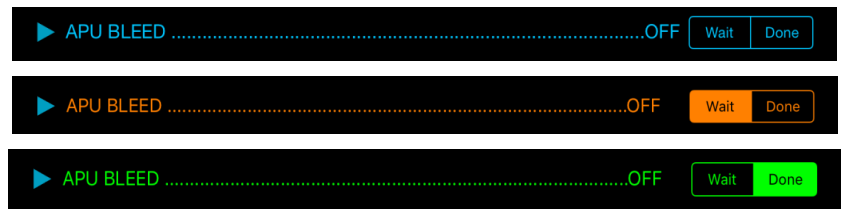

**Fig. 7.** Cyan, green, amber, and white for action status

"Scenarios of human-computer interaction help us to understand and to create computer systems and applications as artifacts of human activity — as things to learn from, as tools to use in one's work, as media for interacting with other people [12]". The current version of OCSIS includes several normal procedures and two abnormal scenarios. "Initial Approach" and "Final approach" are the scenarios that we choose for normal procedures. "Fuel Leak" and "Flaps Locked" are the scenarios that we choose for abnormal procedures. Context patterns trigger procedures in real-time both in normal and abnormal situations. In an abnormal situation such as "Flaps Locked," OCSIS will immediately inform the pilot about this malfunction by displaying a pop-up information window (see Fig. 8). Pilots can become aware of the problem through the pop-up window and start following actions. If they choose to do it later, a reminder line (see Fig. 9) will be displayed at the bottom of the interface, which directs to additional "Flaps Locked" procedures.

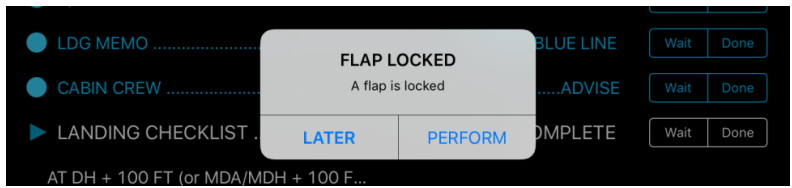

**Fig. 8.** "Flaps Locked" triggering

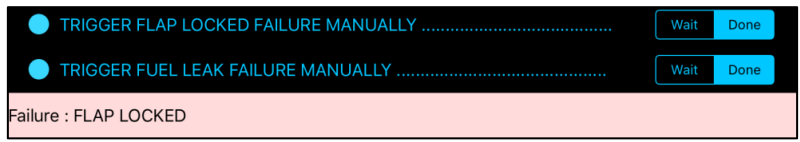

**Fig. 9.** "Flaps Locked" reminder

## 4      Formative evaluation

OCSIS testing complies with the four key types of human factors defined by Chanda and Mongold [13]: (1) Usability of hardware user interface (i.e., we used questionnaires to find out about the location of the OCSIS iPad in the cockpit); (2) Usability of software user interface (i.e., we used questionnaires to get feedback from pilots on usability and usefulness of interface items); (3) Integration of hardware and software with existing cockpit systems (i.e., pilots were asked to provide their opinions on the operational integration of OCSIS in the cockpit); (4) Design of training/procedures for OCSIS (i.e., we designed normal and abnormal procedures for second and third test).

The first testing was carried out at HCDi's simulator lab. Four pilots with flight experience were chosen as participants. They were involved in two sessions. The first session consisted of performing all required procedures both in normal and abnormal scenarios using paper-based manuals. The second session consisted of performing the same procedures using OCSIS. There was an interval of a few days between the two sessions for each person.

Pilot participants provided excellent feedback on actions to use OCSIS, look and feel, and other usability criteria. The results showed that every pilot understood how to use OCSIS. They all reported that OCSIS was easy to hold and use (see Fig. 10). Pilots provided feedback on information icons, color, and size, which was used to improve user interaction with OCSIS. Results showed that all pilots, except one (who said that size of the items was not big enough), felt comfortable with OCSIS information display and interaction (see Fig. 11).

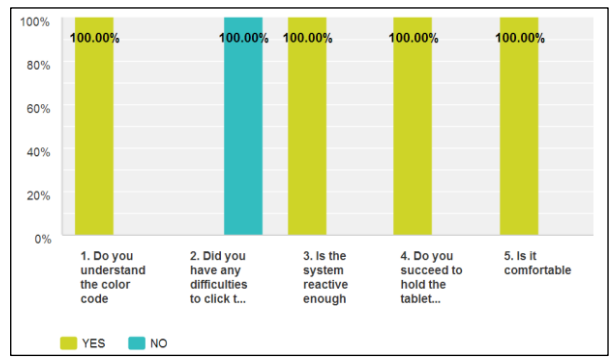

**Fig. 10.** Look and feel feedback

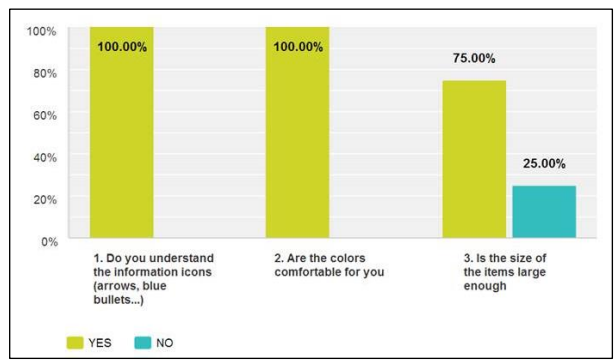

**Fig. 11.** Interaction feedback

Based on pilots' feedbacks we made improvements on part of the icons' color and buttons' function to OCSIS, also changed flight phase titles' size, position and function to remove the ambiguity. A grey color code was added to the system representing the "Not Applicable" case to assist pilots' decision-making. There were procedures that are embedded into other procedures that cause recursivity issues, e.g., the Engine 1(2) Relight procedure is embedded into the "Fuel Leak" procedure. During this phase of testing, we simply added the content of the Engine 1(2) Relight procedure into the "Fuel Leak" part, but we subsequently developed a generic hyperlinked iBlock system within OCSIS that enables to automatically put into an operational sequence.

An integration survey was administered to the pilots. Results showed that all pilots prefer that the iPad be fixed in the cockpit rather than available as a handheld device. Four options were suggested (see Fig. 12): (1) Next to side-stick for each pilot; (2) On the windshield with a flexible arm; (3) In the pedestal or on the instrument panel as a unit; (4) Inside a box at the side of the pedestal.

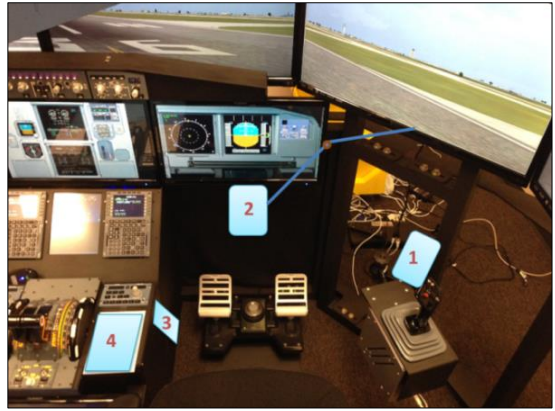

**Fig. 12.** Options of iPad's location

The second testing was carried out at a flight training center in China. Twenty-two A320 pilots, including eleven captains and eleven first officers, participated in the testing, performing as aircrew in A320 simulators. We used the same route and protocols as the first testing. During the testing, we observed in the "Fuel Leak" scenario pilots easily established the failure and excluded the irrelevant procedures (e.g., "Fuel imbalance" procedure). In the "Flaps Locked" scenario, OCSIS reduced the chance of wrong calculation of landing distance and approach speed. Regarding OCSIS look and feel, pilots were asked to evaluate OCSIS usability in terms of colors, interactive buttons, and other devices available on the OCSIS iPad. Results showed that three pilots had difficulty during training to select buttons (see Fig. 13).

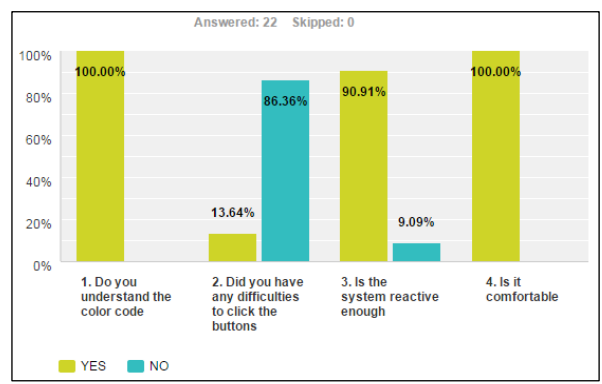

**Fig. 13.** Look and feel feedback

Two pilots thought OCSIS was not reactive enough; this was due to the fact that data-link was not available on the simulator at that time. We corrected that. Pilot-OCSIS interaction was evaluated using pilots' feedback on information icons, color, and size (see Fig. 14). The results showed that pilot-OCSIS interaction is satisfactory in general.

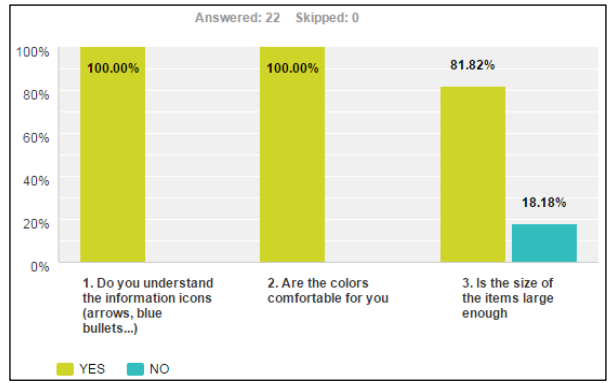

**Fig. 14.** Interaction feedback

Based on pilots' feedback, we made improvements to OCSIS after second testing: "Engine Shutdown" and "Engine Relight" procedures were included in the "Fuel Leak" scenario. Pilots may shut down engine to check if fuel leak is from engine or somewhere else. When pilots move the thrust lever to idle position, the low-speed rotor (N1) of engine is going to be 0, which triggers the "Engine Failure" procedure both on ECAM and OCSIS that increases redundancy. Pilots should perform ECAM actions first and then come back to OCSIS to complete additional actions of "Engine Shutdown" procedure. This being done, pilots can continue executing the "Fuel Leak" procedure, and pilots had to remember to go back procedure. Pilots may decide to relight the engine if they check and discover that the fuel leak is not coming from the engine but from elsewhere. The "Engine Relight" procedure is provided on the same page (see Fig. 15).

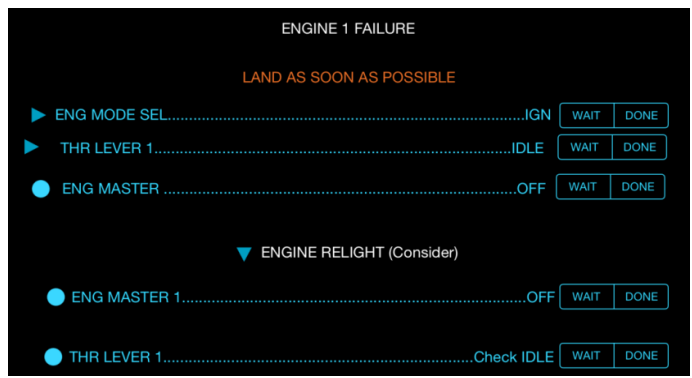

**Fig. 15.** Engine Relight procedure.

Fourteen pilots preferred the iPad located close to each side-stick (Position 1 in Fig. 12). Eight pilots preferred the iPad installed in the middle of instrument panel (Position 2 in Fig. 12). We chose Position 1 as the test location of OCSIS in the third testing. A flexible arm was set up to hold the iPad near the side-stick on the right side of the simulator.

The third testing was held with six pilots participated, and three additional pilots took the Nielsen's ten Usability Heuristics survey [16], using the same protocols and timelines as first and second testing. The results of user-system interaction questionnaires showed that every pilot understood and was satisfied with using OCSIS and with OCSIS's user interaction, as well as the location of OCSIS in the cockpit. Pilots were required to assess OCSIS look and feel by evaluating display usability in terms of color, buttons, and other OCSIS devices. The results show that all pilots had no trouble to understand the system (see Fig. 16). Pilot-OCSIS interaction was assessed on the basis of pilots' feedback on information icons, color, and size (see Fig. 17).

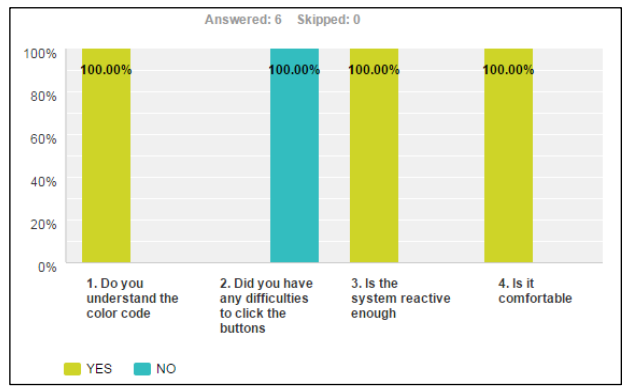

**Fig. 16.** Look and feel feedback

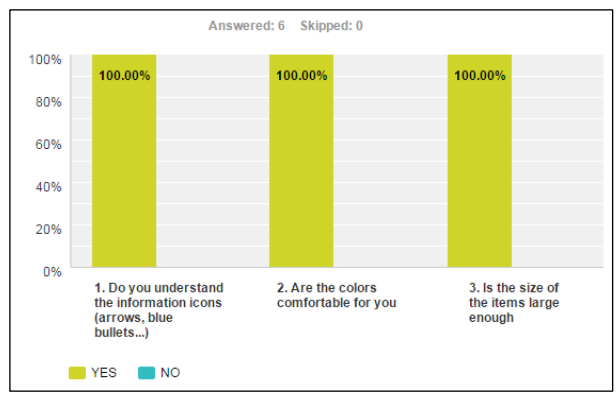

**Fig. 17.** Interaction feedback

Based on Nielsen's ten Usability Heuristics survey [16] which helped us prioritize issues with respect to those that users found critical to those that may not be critical, we made the following improvements to OCSIS after third testing mainly on design decision phase:

1. Quick maneuverability to specific procedures/menus instead of scrolling (see Fig. 18 and Fig. 19). It is possible to fit each flight phase on a single page, and pilots can move left and right to review procedures for other flight phases. In the meantime, a menu to select a particular flight phase can be set at the top of the screen. Pilots should be free to move through menus; information flow progress is saved on each page.

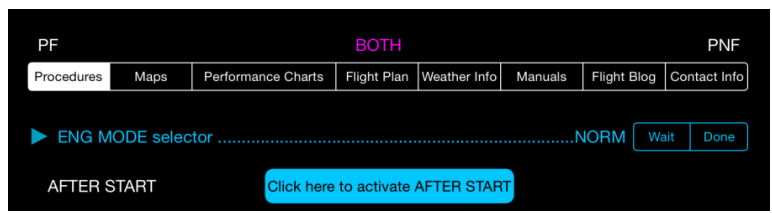

**Fig. 18.** Procedure's headings before improvement

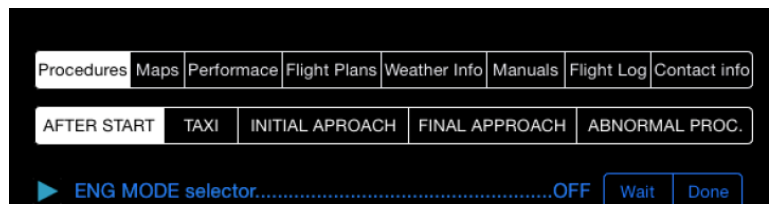

**Fig. 19.** Procedure's headings after improvement

2. Consistency is a key aspect in usability engineering. For example, solid clickable boxes such as "Check-all" bars are inconsistent with the usual cockpit format (e.g., the background of other clickable boxes is black and characters are blue, and as shown on Fig. 20. the solid clickable box format of Check-all bar is the opposite. Fig. 21 is showing the improvement on this point).

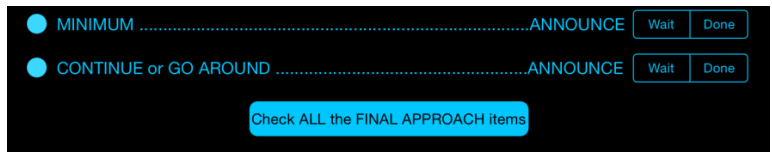

**Fig. 20.** Checklist's icon before improvement

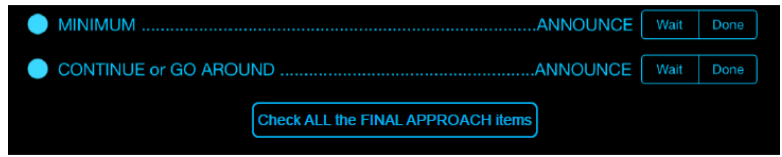

**Fig. 21.** Checklist's icon after improvement

3. Caution or warning messages as well as titles of abnormal procedures should be color coded. Moreover, abnormal procedure headings are not very obvious for every section/page/title of an abnormal procedure. It should be larger (e.g., white "Fuel Leak" title in Fig. 22. It is modified to be amber in Fig. 23).

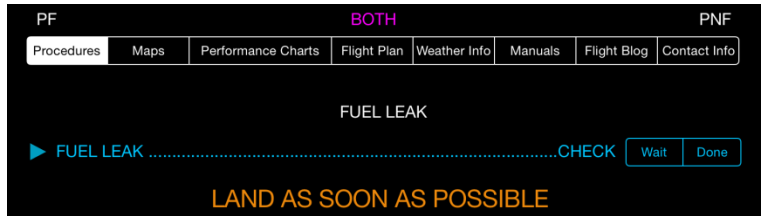

**Fig. 22.** The title of abnormal procedure before improvement

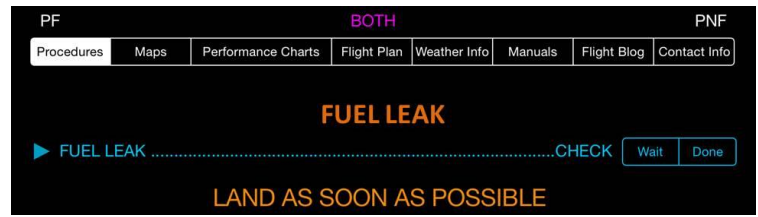

**Fig. 23.** The title of abnormal procedure after improvement

## 5 Discussion

OCSIS was first designed to be used as a tangible interactive system (TIS) onboard a commercial aircraft [17]. Onboard paper-based documentation has been used from the beginning of aviation history and is tangible for pilots to use. Tablets and some applications (e.g., Jeppesen Mobile FliteDeck) have been authorized to be used on the flight deck by the FAA. Consequently, we considered that tablets are tangible objects that can support OCSIS software. This is physical tangibility, but figurative tangibility should be tested [18]. Figurative tangibility in this case means for pilots to keep correct cognition using OCSIS. The testing studies were conducted to provide a first set of methods and tools with this figurative tangibility assessment. And it is possible for us to extend this assessment. The testing studies that were performed showed that this hypothesis was confirmed on a fully equipped cockpit simulator in realistic flight operations scenarios with professional pilots. It is obvious that more iteration is needed and will be implemented in the near future to get a mature version of OCSIS.

This paper provides a first iteration of participatory design of OCSIS. More generally, it shows the shift from the traditional automation approach, where additional software was added to the cockpit and induced some kinds of rigidity that sometimes resulted in unexpected situations, to TIS design, where tangibility has to be tested using situation awareness models and criteria [19]. Of course, the concept of tangibility is more complex and will require more investigation.

## 6　Conclusion

This research and HCD effort are based on both participatory design and agile development (i.e., at the end of each phase, the system is testable in an HITLS environment). This is now typical for the design and development of tangible interactive objects [7], and more generally tangible interactive systems (TISs) [18], where the problem is no longer automation but the search for tangibility. Modeling and simulation are very useful to explore possibilities and drawbacks of these TISs. The quality of both simulation capabilities and pilot participants is crucial [8]. If the issues traditionally raised by human factors and ergonomics specialists when engineering work is done are now posed at the beginning of the design phase in a virtual world (i.e., virtual engineering is part of HCD), new kinds of questions would emerge from this practice, that is, tangibility [20].

OCSIS is a comprehensive system that aims to make flight of commercial aircraft safe, efficient, and comfortable. The multiple usability evaluations and user-centered assessments performed on the system can discover the maximum number of issues. First designs of OCSIS were based on our creativity process, in the sense of synthesis and integration, on previous expertise and experience in the commercial aviation domain, more specifically, work done by Blomberg, Speyer and Boy [11] on the three layers of electronic operations documentation and Ramu's [21] dissertation work on onboard context-sensitive information systems. This work is typically based on human-computer interaction (HCI), hypertext, and context-sensitive information systems work. Usability testing brought us a series of usability issues that helped us to concretely improve OCSIS. Although not all the solutions could be addressed due to time, these can be addressed in further work.

## Acknowledgments

We thank Sebastien Boulnois and Anouk Mirale, who developed a fair amount of OCSIS software, and Varun Korgaonkar who designed the usability questionnaires for the third testing. Thanks to our HCDi colleagues, Jarrett Clark, provided software development support for this project. Dr. Alexandre Lucas Stephane, helped me with the design of testing and the analysis of experimental results, and Dr. Ondrej Doule, who helped me with the design theory and methods. Thanks to all who participated in the various tests and provided me with their time and great feedback to improve OCSIS. Thanks to Dr. Barbara K. Burian, Dr. Divya Chandra, Dr. Christophe Kolski, and Dr. Scott Winter, who provided me with feedback, suggestions, and comments on this project.